\documentclass[11pt]{article}
\usepackage{amsfonts}
\usepackage{amsmath,amssymb,amscd}
\usepackage{stmaryrd}
\usepackage[dvips]{color}
\usepackage{fancybox}
\usepackage{bm}
\usepackage[all]{xy}
\usepackage{wrapfig}

\baselineskip=\normalbaselineskip

\def\Today{\ifcase\month\or January\or February\or March\or April\or May\or
 June\or July\or August\or September\or October\or November\or
December\fi\space\number\day, \number\year(\number\time)}

\def\TM{\widetilde{M}}

\def\condition#1
{\medskip\noindent{\bf Condition\ } #1 :\ }

\def\and{\quad{\rm and}\quad}

\def\p{\partial}

\def\mat#1{\left(\matrix{#1}\right)}

\def\IR{\relax{\rm I\kern-.18em R}}

\def\pr0#1#2#3{{\it Phys.\ Rev.} {\bf #1} (#2) #3}

\def\complex{{\mathchoice
{\setbox0=\hbox{$\displaystyle\rm C$}\hbox{\hbox to0pt
{\kern0.4\wd0\vrule height0.9\ht0\hss}\box0}}
{\setbox0=\hbox{$\textstyle\rm C$}\hbox{\hbox to0pt
{\kern0.4\wd0\vrule height0.9\ht0\hss}\box0}}
{\setbox0=\hbox{$\scriptstyle\rm C$}\hbox{\hbox to0pt
{\kern0.4\wd0\vrule height0.9\ht0\hss}\box0}}
{\setbox0=\hbox{$\scriptscriptstyle\rm C$}\hbox{\hbox to0pt
{\kern0.4\wd0\vrule height0.9\ht0\hss}\box0}}}}
\def\Co{{\mathchoice
{\setbox0=\hbox{$\displaystyle\rm C$}\hbox{\hbox to0pt
{\kern0.4\wd0\vrule height0.9\ht0\hss}\box0}}
{\setbox0=\hbox{$\textstyle\rm C$}\hbox{\hbox to0pt
{\kern0.4\wd0\vrule height0.9\ht0\hss}\box0}}
{\setbox0=\hbox{$\scriptstyle\rm C$}\hbox{\hbox to0pt
{\kern0.4\wd0\vrule height0.9\ht0\hss}\box0}}
{\setbox0=\hbox{$\scriptscriptstyle\rm C$}\hbox{\hbox to0pt
{\kern0.4\wd0\vrule height0.9\ht0\hss}\box0}}}}
\def\Rl{{\mathchoice
{\setbox0=\hbox{$\displaystyle\rm R$}\hbox{\hbox to0pt
{\kern0.4\wd0\vrule height0.9\ht0\hss}\box0}}
{\setbox0=\hbox{$\textstyle\rm R$}\hbox{\hbox to0pt
{\kern0.4\wd0\vrule height0.9\ht0\hss}\box0}}
{\setbox0=\hbox{$\scriptstyle\rm R$}\hbox{\hbox to0pt
{\kern0.4\wd0\vrule height0.9\ht0\hss}\box0}}
{\setbox0=\hbox{$\scriptscriptstyle\rm R$}\hbox{\hbox to0pt
{\kern0.4\wd0\vrule height0.9\ht0\hss}\box0}}}}

\begin{document}

\numberwithin{equation}{section}
\renewcommand{\theequation}{\thesection.\arabic{equation}}
\def\natural{\mathbb{N}}
\def\mat#1{\matt[#1]}
\def\matt[#1,#2,#3,#4]{\left(%
\begin{array}{cc} #1 & #2 \\ #3 & #4 \end{array} \right)}
\def\hq{\hat{q}}
\def\hp{\hat{p}}
\def\hx{\hat{x}}
\def\hk{\hat{k}}
\def\hw{\hat{w}}
\def\hl{\hat{l}}

\def\bea#1\ena{\begin{align}#1\end{align}}
\def\nn{\nonumber\\}
\def\cL{{\cal L}}
\def\TM{TM\oplus T^*M}
\newcommand{\CouB}[2]{\left\llbracket #1,#2 \right\rrbracket}
\newcommand{\pair}[2]{\left\langle\, #1, #2\,\right\rangle}

\null \hfill Preprint TU-978  \\[3em]
\begin{center}
{\LARGE \bf{
Poisson-generalized geometry and $R$-flux
}}
\end{center}

\begin{center}
{T. Asakawa${}^\sharp$\footnote{
e-mail: asakawa@maebashi-it.ac.jp}, H. Muraki${}^{\flat}$\footnote{
e-mail: hmuraki@tuhep.phys.tohoku.ac.jp}, S. Sasa${}^{\flat}$\footnote{
e-mail: sasa@tuhep.phys.tohoku.ac.jp} and S. Watamura${}^{\flat}$\footnote{
e-mail: watamura@tuhep.phys.tohoku.ac.jp}}\\[3em] 
${}^\sharp$
Department of Integrated Design Engineering,\\
Faculty of Engineering,\\
Maebashi Institute of Technology\\
Maebashi, 371-0816, Japan \\[1em]

${}^\flat$
Particle Theory and Cosmology Group \\
Department of Physics \\
Graduate School of Science \\
Tohoku University \\
Aoba-ku, Sendai 980-8578, Japan \\ [5ex]

\thispagestyle{empty}

\abstract{
We study a new kind of Courant algebroid on Poisson manifolds, 
which is a variant of the generalized tangent bundle
in the sense that the roles of tangent and the cotangent bundle are exchanged.
Its symmetry is a semidirect product of $\beta$-diffeomorphisms and $\beta$-transformations.
It is a starting point 
of an alternative version of the generalized geometry based on the cotangent bundle, 
such as Dirac structures and generalized Riemannian structures.
In particular, $R$-fluxes are formulated as a twisting of this Courant algebroid by a local $\beta$-transformations,  
in the same way as $H$-fluxes are the twist of the generalized tangent bundle.
It is a $3$-vector classified by Poisson $3$-cohomology and it appears in a twisted bracket 
and in an exact sequence.} 

\end{center}

\vskip 2cm

\eject
\section{Introduction}

The effective theory of string is given by
the $10$-dimensional supergravity coupled to matter fields.
In the supergravity we encounter various kinds of fluxes.
Among them the NS-NS $B$-field is of particular interest.
Since the $B$-field appears as the antisymmetric counterpart of the metric $g$,
it is natural to consider $g$ and $B$ on the same footing in the stringy geometry.
Furthermore, since they are mixed by the T-duality transformation,
they should be unified.
By using the generalized geometry \cite{Hitchin,Gualtieri},
we can realize a formulation along this line
where T-duality is manifest \cite{Bouwknegt:2003zg,Cavalcanti:2011wu}.

It is known that the T-duality transformations of the theory with non-trivial $H$-flux
give rise to exotic spaces accompanied with fluxes of new types.
Such a space is referred to as ``non-geometric space" in literatures,
though the geometrical meaning of the non-geometric space is still unclear.
The stringy geometry should explain at least the meaning of those new types of fluxes. 
In this paper, we want to propose a variant of the generalized geometry, 
which will help to formulate the space with those various types of fluxes.

Non-geometric fluxes are first recognized in the study of $4$-dimensional gauged supergravity.
They appear in the Kaloper-Myers algebra \cite{Kaloper}
\begin{align*}
&[e_a, e_b]=f_{ab}^c e_c + H_{abc}e^c,\\
&[e_a, e^b]=Q_a^{bc} e_c + f_{ac}^b e^c,\\
&[e^a, e^b]=R^{abc} e_c + Q_c^{ab}e^c.
\end{align*}
Here $H_{abc}$ is the $H$-flux, $f_{ab}^c$ is a geometric-flux, $Q_a^{ab}$ and $R^{abc}$ are 
so-called non-geometric fluxes.
T-duality brings also such fluxes into discussion.
In \cite{Hull:2004in,Shelton:2006fd} the authors investigate, as
an example, T-duality sequence of a $3$-dimensional torus $T^3$ with constant $H$-flux.
Since there are three isometries (translation along each direction), 
it is suggested that T-duality transformation can be applied three times,
which relates $H$-flux and others as 
\begin{align*}
H_{abc} \longrightarrow f_{ab}^c \longrightarrow Q_a^{bc} \longrightarrow R^{abc}.
\end{align*}
In this example the spaces with $Q$-flux is locally geometric, but globally non-geometric.
The third T-duality transformation is not really understood,
and the geometrical meaning of the space with $R$-flux is rather unclear. 
However there are arguments \cite{Dabholkar:2005ve, Grana} which strongly suggest
the existence of non-geometric fluxes in superstring theory.

Theories in such non-geometric flux backgrounds are investigated from various viewpoints.
String worldsheet theories in such backgrounds are studied 
\cite{Halmagyi:2008dr}, which originate from the work by \cite{Alekseev:2004np}
for review see \cite{Zabzine}.
An interesting observation is made by \cite{Blumenhagen}
that the $R$-flux makes spacetime non-commutative and non-associative.
From the target spacetime point of view, a $10$-dimensional supergravity 
with non-geometric fluxes is formulated \cite{Blumenhagen:2012nt, Andriot:2013xca}.
There the notions of $\beta$-diffeomorphisms and $\beta$-tensors are introduced,
and the $R$-flux is identified with the violation of the Poisson structure.
These papers and \cite{Chatzistavrakidis:2013wra}
suggest the importance of the use of (quasi-)Lie algebroid of a Poisson manifold 
$(T^*M)_\theta$ or its variant in order to formulate $R$-flux.
However, the proper treatments and the properties
of non-geometric fluxes are still mysterious.

The aim of this paper is to propose a natural definition of $R$-flux.
To define it, we demand the following properties
expected from the previous studies: 
It should be a $3$-vector defined globally; and
its underlying symmetry should be based on $\beta$-transformations.
We do not use any properties of the fluxes under T-duality.
The reason is as follows.  
The T-duality assumes the existence of an isometry
and thus a (generalized) Riemannian structure.
However in the formulation of $H$-flux these additional assumptions are not required.
Hence it is natural to consider that there is no such type of assumption
 in the formulation of $R$-flux.
In this sense, our definition of the $R$-flux is not a derivation but a proposal, 
and its validity in the context of superstring theory should be analyzed separately.

It is known that the $H$-flux is  naturally introduced
 within the framework of the generalized geometry.
The generalized tangent bundle $TM\oplus T^*M$ has the symmetry of $B$-transformations.
The $H$-flux is used for a twist to define an exact Courant algebroid $E$ satisfying
\begin{align}
0\to T^*M \to E \to TM \to 0.
\end{align}
Here $E$ is glued by local $B$-gauge transformations
and the corresponding flux is the $H$-flux.

In this paper, to formulate the $R$-flux, we
imitate this formulation of the $H$-flux.
We first introduce a new Courant algebroid $(TM)_0\oplus (T^*M)_\theta$ on a Poisson manifold,
which has the symmetry of $\beta$-transformations.
There the roles of the tangent bundle $TM$ and the cotangent bundle $T^*M$ are exchanged
compared to the standard generalized tangent bundle $TM\oplus T^*M$.
Then a $3$-vector flux $R$ is used for the twisting, 
which leads to an exact sequence
\begin{align}
0\to (TM)_0 \to E \to (T^*M)_\theta \to 0.
\end{align}
Here $E$ is glued by local $\beta$-gauge transformations and the corresponding flux is the $R$-flux.
The point in this paper is to use the different Courant algebroid 
from the one used in the formulation of the $H$-flux.

Note that in our definition of $R$-flux, the underlying structure is not the quasi-Poisson structure
but the Poisson structure. And the $R$-flux is
identified with a flux given by a bivector field ``gauge potential'', 
in the similar manner as the $H$-flux which is given by a 2-form gauge potential.

The Courant algebroid $TM\oplus T^*M$ 
is the starting point of the standard generalized geometry, where
the various concepts are developed such as
 the Dirac, generalized complex, generalized Riemannian structures
and etc..
They are considered as the unified objects
 of various structures appearing in the ordinary differential geometry, which is
based on the tangent bundle $TM$.

Correspondingly, our new Courant algebroid $(TM)_0\oplus (T^*M)_\theta$ gives the 
base of the alternative of the generalized geometry, 
that we call the Poisson-generalized geometry in this paper.
It should be an extension of the Poisson geometry, 
the differential geometry based on $(T^*M)_\theta$.
In this paper we give some preliminary discussion to this direction.

The plan of this paper is as follows.
In section 2, 
we introduce the 
Courant algebroid $(TM)_0\oplus (T^*M)_\theta$ and discuss its properties.
We find that its symmetry
consists of $\beta$-diffeomorphisms and $\beta$-transformations.
In section 3,
we propose a definition of the $R$-fluxes based on local $\beta$-transformations.
We see that
the mathematical structure of the $R$-fluxes is quite similar to that of the $H$-fluxes.
After that, in section 4 we give some preliminary results on the Poisson-generalized geometry.
We discuss about Dirac structures, generalized Riemannain structures, and pure spinors.
Final section is devoted to conclusion and discussion about future direction.

\section{New Courant algebroid}

In this section, after recalling some basic definitions of the Lie algebroid $(T^*M)_\theta$
of a Poisson manifold,
we introduce a new Courant algebroid $(TM)_0\oplus (T^*M)_\theta$. 
The corresponding bracket is 
different from the Courant bracket of the Courant algebroid $TM\oplus T^*M$,
which is used in the standard generalized geometry.
Here we investigate the symmetry of the new Courant algebroid.

\subsection{Lie algebroid of a Poisson manifold}

Let $(M,\theta)$ be a Poisson manifold equipped with a Poisson bivector $\theta \in \wedge^2 TM$.
The Poisson bivector $\theta$ satisfies $[\theta,\theta]_S=0$, 
where $[\cdot,\cdot]_S$ is the Schouten-Nijenhuis bracket.
A Lie algebroid of a Poisson manifold \cite{CdSWeinstein} is defined 
by a triple $(T^*M, \, \theta, \, [\cdot,\cdot]_\theta)$,
where $T^*M$ is the cotangent bundle;
$\theta$ is an anchor map, which is obtained by regarding the Poisson bivector $\theta$
as a map $\theta :T^*M\to TM$, i.e. $\theta(\xi)=i_\xi \theta$ for $\xi \in T^*M$;
and a Lie bracket $[\cdot,\cdot]_\theta$ is  defined by the Koszul bracket:
\bea
[\xi,\eta]_\theta ={\cal L}_{\theta (\xi)}\eta-i_{\theta (\eta)}d\xi .
\ena
We also denote this Lie algebroid $(T^*M,\,  \theta, \, [\cdot,\cdot]_\theta)$ 
as $(T^*M)_\theta$ for short.

In general, a Lie algebroid $A$ defines 
a differential algebra $(\Gamma (\wedge^\bullet A^*),\, \wedge, \, d_A)$ of $A$-forms
and a Gersternhaber algebra $(\Gamma (\wedge^\bullet A),\, \wedge, \, [\cdot,\cdot]_A)$
of $A$-polyvectors, where $d_A$ is the $A$-exterior derivative and
$[\cdot,\cdot]_A$ is the Gersternhaber bracket, extension of the Lie bracket \cite{CdSWeinstein}.
From a Lie algebroid, $A$-Lie derivative acting on $\Gamma (\wedge^\bullet A^*)$ 
as well as $\Gamma (\wedge^\bullet A)$ is defined and satisfies the $A$-Cartan relation.

In our case with $A=(T^*M)_\theta$,
we can define the corresponding objects as follows:
$\Gamma (\wedge^\bullet A^*)=\Gamma (\wedge^\bullet TM)$ is the space of polyvectors, 
and the exterior derivative is $d_A=d_\theta =[\theta, \cdot]_S $.
In particular, for a function $f\in C^\infty (M)$, it acts as
\bea
d_\theta f =[\theta, f]_S =-\theta (df).
\ena
The actions of the Lie derivative ${\cal L}_{\zeta}$, where $\zeta \in T^*M$, 
on  a function $f$, a $1$-form $\xi$ and a vector field $X$ are given by
\bea
&{\cal L}_{\zeta} f :=i_\zeta d_\theta f, \nn
&{\cal L}_{\zeta} \xi :=[\zeta,\xi]_\theta , \nn
&{\cal L}_{\zeta} X :=(d_\theta i_\zeta +i_\zeta d_\theta) X,
\label{A Lie derivative}
\ena
respectively.
The corresponding Cartan relation on the space of polyvectors $\Gamma( \wedge^\bullet TM)$ is
\bea
\{ i_\xi, i_\eta\}=0, \quad
\{ d_\theta ,i_\xi\}={\cal L}_\xi, \quad
[{\cal L}_\xi, i_\eta ]=i_{[\xi,\eta]_\theta},\quad
[{\cal L}_\xi, {\cal L}_\eta ]={\cal L}_{[\xi,\eta]_\theta},\quad
[d_\theta,{\cal L}_\xi ]=0.
\label{A Cartan}
\ena

\subsection{Courant algebroid $(TM)_0\oplus (T^*M)_\theta$}

Consider a vector bundle $TM\oplus T^*M$ with a canonical inner product 
\bea
\langle X+\xi,Y+\eta\rangle =\textstyle{\frac{1}{2}}(i_X \eta +i_Y \xi),
\ena
an anchor map $\rho :TM\oplus T^*M \to TM$ given by
\bea
\rho(X+\xi)=\theta (\xi),\label{Sasa anchor}
\ena
and a skew-symmetric bracket 
\bea
[X+\xi,Y+\eta ] =[\xi,\eta ]_\theta +{\cal L}_{\xi}Y -{\cal L}_{\eta}X 
+\frac{1}{2}d_\theta (i_X\eta -i_Y\xi ).
\label{Sasa bracket}
\ena
Then the quadruple $(TM\oplus T^*M, \langle \cdot,\cdot \rangle,\rho,[\cdot,\cdot])$ 
is a Courant algebroid.
To show this, note that a Lie bialgebroid $A\oplus A^*$ is always a Courant algebroid \cite{LWX},
and the above Courant algebroid is of this type.
Here the first Lie algebroid $(TM)_0:=(TM,\, a=0,\, [\cdot,\cdot]=0)$ 
is a tangent bundle with the vanishing Lie bracket and the vanishing anchor map,
while the second $(T^*M)_\theta =(T^*M, \, \theta, \, [\cdot,\cdot]_\theta)$ 
is the Lie algebroid of a Poisson manifold explained in the previous subsection 2.1.
We denote this Courant algebroid $(TM\oplus T^*M, \langle \cdot,\cdot \rangle,\rho,[\cdot,\cdot])$
 as $(TM)_0\oplus (T^*M)_\theta$.

To get some ideas of the Courant algebroid $(TM)_0\oplus (T^*M)_\theta$,
it is instructive to recall some notions of
the standard generalized tangent bundle $TM\oplus T^*M$.
In the standard case, the anchor map is given by
$\rho (X+\xi)=X$ and the Courant bracket 
$[\cdot,\cdot]_C$ is 
\bea
[X+\xi,Y+\eta ]_C
&=[X,Y]+{\cal L}_X \eta-{\cal L}_Y \xi -\frac{1}{2}d (i_X\eta -i_Y\xi ).
\label{Courant bracket}
\ena
The Courant algebroid  $TM\oplus T^*M$
can be considered as an extension of the Lie algebroid $TM$,
and in fact it is a Lie bialgebroid 
$(TM, \,{\rm id.},\,[\cdot,\cdot]_{TM})\oplus(T^*M, \,0, \,0)$.

Contrary to this, in our Courant algebroid,
the roles of $TM$ and $T^*M$ are exchanged:
The underlying Lie bialgebroids is $(T^*M, \, \theta, \, [\cdot,\cdot]_\theta)\oplus(TM,\, 0,\, 0)$; 
the anchor map (\ref{Sasa anchor}) picks up only $T^*M$-part;
the Courant bracket (\ref{Sasa bracket}) 
is written in terms of the operations defined in $(T^*M)_\theta$ only.
In this way, our Courant algebroid $(TM)_0\oplus (T^*M)_\theta$ 
can be considered as an extension of  
the Lie algebroid $(T^*M)_\theta$.

As a consequence,
in the Courant algebroid $(TM)_0\oplus (T^*M)_\theta$,
the Poisson Lie algebroid $(T^*M)_\theta$ governs the differential geometry, 
and the resulting differential geometry is quite different from the one
governed by the Lie algebroid $TM$
as seen from the structure function of $(T^*M)_\theta$.
However, we can proceed to formulate an analogue of 
the generalized geometry exactly in the same manner as
the standard generalized tangent bundle.

Some comments are in order:
First, the standard Courant bracket (\ref{Courant bracket})
and the new bracket (\ref{Sasa bracket}) can be considered as 
complementary parts in the Roytenberg bracket \cite{Halmagyi:2008dr,
Roytenberg:2001am}:
\bea
[X+\xi,Y+\eta ]_{\rm Roy} 
&=[X,Y]+{\cal L}_X \eta-{\cal L}_Y \xi -\frac{1}{2}d (i_X\eta -i_Y\xi )\nn
&+[\xi,\eta ]_\theta +{\cal L}_{\xi}Y -{\cal L}_{\eta}X 
+\frac{1}{2}d_\theta (i_X\eta -i_Y\xi ).
\label{Roytenberg bracket}
\ena
Note that the Roytenberg bracket is the bracket
for a Lie bialgebroid $TM\oplus (T^*M)_\theta$
and not for the present Courant algebroid $(TM)_0\oplus (T^*M)_\theta$.

Secondly, in general, an anchor map $\rho:E\to TM$ of a Courant algebroid $E$ 
induces a natural differential operator 
$D:C^\infty (M) \to \Gamma(E)$ defined by
$\langle Df,A\rangle=\textstyle{\frac{1}{2}}\rho (A)\cdot f,$
for arbitrary function $f \in C^\infty (M)$ and section $A \in \Gamma(E)$.
In our case, 
\bea
\langle Df,X+\xi\rangle=\textstyle{\frac{1}{2}}\theta(\xi)\cdot f 
=\textstyle{\frac{1}{2}}\theta (\xi,df) =-\textstyle{\frac{1}{2}} i_\xi \theta (df),
\ena
implies that $Df=d_\theta f =-\theta (df) \in \Gamma(TM)$.
This also follows from the general construction of $D=d_A+d_{A^*}$ 
in the Lie bialgebroid $A\oplus A^*$.
Here $A=(TM)_0$ and thus $d_A=0$ and $A^*=(T^*M)_\theta$ and thus $d_{A^*}=d_\theta$.

Finally,
In \cite{AMW}, the same Lie algebroid $(T^*M)_\theta$ is used but in a different context. 
It appears as a Dirac structure in the standard generalized tangent bundle $TM\oplus T^*M$.

\subsection{Symmetry of $(TM)_0\oplus(T^*M)_\theta$}

It is known that the symmetry of the generalized tangent bundle $TM\oplus T^*M$ is the 
semidirect product of diffeomorphisms generated by vector fields,  and $B$-transformations 
with closed $2$-forms.
Here we investigate the symmetry of our Courant algebroid $(TM)_0\oplus(T^*M)_\theta$.

Let us define the 
following two transformations acting on a section $X+\xi \in (TM)_0\oplus(T^*M)_\theta$:
\begin{enumerate}
\item $\beta$-diffeomorphism:
For a $1$-form $\zeta \in T^*M$, we define
\bea
{\cal L}_{\zeta}(X+\xi)={\cal L}_{\zeta}X+ {\cal L}_{\zeta}\xi,
\ena
by the diagonal action of the Lie derivative ${\cal L}_{\zeta}$ given in (\ref{A Lie derivative}).
\item $\beta$-transformation: For a bivector $\beta \in \wedge^2 TM$, we define
\bea
e^{\beta}(X+\xi)= X+ \xi +i_\xi \beta.
\ena
\end{enumerate}
The $\beta$-transformation is a widely-used definition 
in the context of $TM\oplus T^*M$ and is not a symmetry of the Courant bracket of $TM\oplus T^*M$.
The $\beta$-diffeomorphism is a natural 
object for  the Courant algebroid $(TM)_0\oplus(T^*M)_\theta$ of a Poisson manifold as follows.
It is instructive to see it from the viewpoint of $TM\oplus T^*M$.
To this end, rewrite (\ref{A Lie derivative}) following \cite{Blumenhagen:2012nt} as
\bea
&{\cal L}_{\zeta} f ={\cal L}_{\theta(\zeta)}f, \notag \\
&{\cal L}_{\zeta} \xi  ={\cal L}_{\theta(\zeta)}\xi -i_{\theta(\xi)}d\zeta,\notag \\
&{\cal L}_{\zeta} X ={\cal L}_{\theta(\zeta)}X +\theta (i_X d\zeta), \label{gege beta-diffeo}
\ena
The third equation of (\ref{gege beta-diffeo}) is proven in the appendix \ref{app beta Lie}.
In the above expressions, the terms of the ordinary Lie derivative ${\cal L}_{\theta(\zeta)}$
represent a diffeomorphism generated by a vector field $\theta(\zeta)$.
The term $i_{\theta(\xi)}d\zeta$ in the second equation
 is a $B$-transformation with $d\zeta$ of a $\beta$-transformed vector $\theta(\xi)$,
while the term $\theta (i_X d\zeta)$ in the third equation is 
a $\beta$-transformation of a $B$-transformation with $d\zeta$.
Therefore, the $\beta$-diffeomorphism is a rather complicated combination of a diffeomorphism, 
a $B$-transformation and a $\beta$-transformation from the viewpoint of $TM\oplus T^*M$.
And it is no longer a symmetry of the Courant bracket of $TM\oplus T^*M$.

It is worth mentioning that
if the parameter $\zeta$ is exact, $\zeta=dh$,
the $\beta$-diffeomorphism (\ref{A Lie derivative})  reduces to the ordinary diffeomorphism
generated by the Hamilton vector field $X_h =\theta(dh)$:
\bea
{\cal L}_{dh}X= {\cal L}_{X_h}X, \quad {\cal L}_{dh}\xi= {\cal L}_{X_h}\xi.
\ena
Such exact $1$-forms form a subgroup of the group of $\beta$-diffeomorphisms.

Note that the name of the $\beta$-diffeomorphism is introduced in \cite{Blumenhagen:2012nt}.
The authors considered the above Lie derivative in the case with the quasi-Poisson structure
$[\theta,\theta]_S\not=0$, and identified this violation of the Poisson structure
with an $R$-flux. 
On the other hand, in this paper
we are considering the case that the bivector $\theta$ is exactly Poisson $[\theta,\theta]_S=0$.
Although we use the same terminology as in \cite{Blumenhagen:2012nt}, 
our proposal for the $R$-flux is associated
not with $\beta$-diffeomorphisms but with $\beta$-transformations, as we propose in the following.

We are now ready to study the symmetry of the new Courant algebroid.
In general, a symmetry of the Courant algebroid $E$ is a bundle map $\varphi :E\to E$ 
such that it is compatible with the three structures, that is, for $A,B \in E$,
\bea
&\langle \varphi A, \varphi B\rangle =\varphi \langle A,B\rangle, \nn
&\rho (\varphi A) =\varphi \rho (A) ,\notag \\
&[\varphi A,\varphi B]=\varphi [A,B].  \label{general symmetry}
\ena
In the case we are considering,
the right hand side of the first and the second equations $\varphi$ denote the induced actions on 
$C^\infty (M)$ and $TM$, respectively.

For an infinitesimal $\beta$-diffeomorphism ${\cal L}_{\zeta}$, the equations
 (\ref{general symmetry}) read
\bea
&\langle {\cal L}_{\zeta} A, B\rangle +\langle A, {\cal L}_{\zeta}B\rangle 
={\cal L}_{\zeta} \langle A,B\rangle, \notag \\
&\rho ({\cal L}_{\zeta} A) ={\cal L}_{\zeta}\rho (B) , \notag \\
&[{\cal L}_{\zeta} A, B]+[A, {\cal L}_{\zeta}B]={\cal L}_{\zeta} [A,B]. \label{sym:beta-diffeo}
\ena
The first and the third equations are satisfied by an arbitrary $\zeta$, while the second equation 
holds when the vector field $\theta (\zeta)$ is $d_\theta$-closed. The proofs of the above relations
are given in the appendix \ref{app beta diffeo}.

For a $\beta$-transformation $e^\beta$, the equations (\ref{general symmetry}) read
\bea
&\langle e^\beta A, e^\beta B\rangle =\langle A,B\rangle, \notag \\
&\rho (e^\beta A) =\rho (B) ,\notag \\
&[e^\beta A, e^\beta B]=e^\beta[A,B]. \label{sym:beta-trf}
\ena
We can show that the first and the second equations are satisfied by an arbitrary $\beta$, 
while the last equation holds when the bivector field $\beta$ is $d_\theta$-closed.
First two equations are obvious to hold.
We give the proof of the third equation in the appendix \ref{app beta transf}.

In summary, a $\beta$-diffeomorphism ${\cal L}_\zeta$ is a symmetry 
if ${\cal L}_\zeta \theta =0$
and a $\beta$-transformation $e^\beta$ is a symmetry if $d_\theta \beta=0$.
This result shows that also in the symmetry structure,
the roles of $TM$ and $T^*M$ are interchanged, 
compared to the standard generalized tangent bundle.
In particular, for constructing $R$-fluxes, 
it is essential that the $\beta$-transformations are the symmetry of the new bracket,
as we shall see in the next section.

We end this section with a few remarks.
As in the case of $B$-transformation, we call the particular case of a $\beta$-transformation 
$e^{d_\theta Z}$ with a $d_\theta$-exact bivector 
$\beta=d_\theta Z$, a $\beta$-gauge transformation.
Similar to the Courant bracket of $TM\oplus T^*M$,
the action of a pair $(\zeta, \beta)=(\zeta, -d_\theta Z)$ can be written as
\bea
{\cal L}_{(\zeta, -d_\theta Z)}(X+\xi)
&=[\zeta,\xi ]_\theta +{\cal L}_{\zeta}X-i_{\xi}d_\theta Z \nn
&=(\zeta +Z) \circ (X+\xi ),
\ena
where in the last line, the symbol $\circ $ denotes the analogue of the Dorfman bracket
\footnote{The skew-symmetrization of $\circ $ gives the new bracket (\ref{Sasa bracket})}.
Hence, a $\beta$-gauge transformation is an inner transformation.

It is also worth to note that
the $\beta$-transformation does not yield a naive shift $\theta \to \theta+\beta$ 
of the bivector $\theta$.
Here the situation is different from the case in the paper \cite{AMW}, where 
$(T^*M)_\theta $ is regarded as a Dirac structure in $TM\oplus T^*M$,
and the $\beta$-transformation is required to  preserve the Dirac structure.
In that case the $\beta$-transformation indeed results in a shift $\theta \to \theta+\beta$,
and the Maurer-Cartan type condition for $\beta$ has to be satisfied.

\section{Proposal of $R$-flux}

In this section, we propose a definition of $R$-fluxes by a set of data $(R,\beta_i,\alpha_{ij})$,
where $R\in \wedge^3 TM$, $\beta_i \in \wedge^2 TU_i$ and $\alpha_{ij} \in TU_{ij}$,
such that
\bea
&R|_{U_i}=d_\theta \beta_i , \nn
&\beta_j -\beta_i|_{U_{ij}} =d_\theta \alpha_{ij}.
\label{R-flux}
\ena
Here $\{U_i\}$ denotes a good open covering of $M$ and $U_{ij}=U_i \cap U_j$.
It follows from (\ref{R-flux}) that $R$ is a global $3$-vector on $M$
and is $d_\theta$-closed: $d_\theta R=0$.
Local bivectors $\{\beta_i\}$ are gauge potentials for the $R$-flux, the analogue of $B$-fields for $H$-fluxes, 
and correspondingly, there is the local $\beta$-gauge symmetry of the form
\bea
\beta_i \mapsto \beta_i +d_\theta \Lambda_i, \quad 
\alpha_{ij} \mapsto \alpha_{ij} +\Lambda_i -\Lambda_j, 
\label{local beta gauge symmetry}
\ena
for an arbitrary gauge parameter $\Lambda_i \in TU_i$.
In particular, since the $R$-flux is invariant under the gauge symmetry, it is abelian.

This proposal is based on the mathematical correspondence between 
the standard generalized tangent bundle 
$TM\oplus T^*M$ and our new Courant algebroid $(TM)_0 \oplus (T^*M)_\theta$.
In the following we show that this $R$-flux is exactly the $(TM)_0 \oplus (T^*M)_\theta$-analogue 
of an $H$-flux in $TM\oplus T^*M$.
Concerning the definition of the $H$-flux, see the appendix \ref{app H-flux}.

Recall that in the new Courant algebroid $(TM)_0 \oplus (T^*M)_\theta$, comparing with $TM\oplus T^*M$,
$(T^*M)_\theta$ play the role of $TM$.
Thus, an $H$-twisting of $TM\oplus T^*M$ (\ref{twisting})
corresponds to a twisting of $(TM)_0 \oplus (T^*M)_\theta$
satisfying the exact sequence
\bea
0\to (TM)_0 \xrightarrow{\pi^*} E \xrightarrow{\pi} (T^*M)_\theta \to 0.
\label{R-exact sequance}
\ena
We emphasize that  the bundle map $\pi$ is {\it not} an anchor map, thus the meaning of 
the exactness of (\ref{R-exact sequance}) is different from the standard exact Courant algebroid.

In the following subsection we show
\begin{enumerate}
\item Given a data $(R,\beta,\alpha)$ in (\ref{R-flux}) 
we can construct a Courant algebroid $E$ that satisfies 
the exact sequence (\ref{R-exact sequance}).
It is classified by Poisson cohomology $[R] \in H^3_\theta (M)$.
\item $E$ is isomorphic to the untwisted Courant algebroid $(TM)_0 \oplus (T^*M)_\theta$ 
with the $R$-twisted bracket.
\item $E$ is a quasi-Lie bialgebroid $((TM)_0, \delta=0, \phi=R)$.
\end{enumerate}
Each statement has its analog in the case of $H$-fluxes \cite{SeveraWeinstein, Rogers}, 
here our logic is near to the one developed by \cite{Rogers}.

\subsection{Gluing by local $\beta$-gauge transformation}

We follow the argument of [21] for H-fluxes, but replace the role
of $TM$ with that of $T^*M$, 
 and $B$-transformations with $\beta$-transformations.

Let $(M,\theta)$ be a $d$-dimensional Poisson manifold with a $d_\theta$-closed 
$3$-vector $R \in \wedge^3 TM$. 
We assume a trivialization of $R$, that is, an open cover $\{U_i\}$ of $M$ equipped with 
local bivectors $\beta_i \in \wedge^2 TU_i$ and vectors $\alpha_{ij} \in TU_{ij}$ such that
(\ref{R-flux}) is satisfied.
Given such a trivialization, a Courant algebroid $E$ is constructed in the following way.
First, over each open set $U_i$, we can consider a Courant algebroid 
$E_i=(TU_i)_0 \oplus (T^*U_i )_\theta $, equipped with the anchor map $\rho_i$, 
the inner product $\langle\cdot,\cdot\rangle_i$ and the bracket $[\cdot,\cdot]_i$ defined by
\bea
&\rho_i (\xi)=\theta (\xi), \quad
\langle X+\xi ,Y+\eta \rangle_i =\textstyle{\frac{1}{2}}(i_X\eta -i_Y\xi),\nn
&[X+\xi, Y+\eta]_i=[\xi,\eta]_\theta +{\cal L}_\xi Y-{\cal L}_\eta X+\textstyle{\frac{1}{2}}d_\theta (i_X\eta -i_Y\xi),
\ena
for $X+\xi, Y+\eta \in (TU_i)_0 \oplus (T^*U_i )_\theta$.
On the intersection $U_{ij}$, $E_i$ and $E_j$
are glued by a $\beta$-gauge transformation generated by $\alpha_{ij}$, that is the transition function is
\bea
&G_{ij} :U_{ij} \to O(d,d),\nn
&G_{ij}(x) =\mat{1,-d_\theta \alpha_{ij}(x),0,1}.
\ena
It defines the equivalence relation $\sim $ between $X+\xi \in E_j|_{U_{ij}}$ and
\bea
G_{ij} (X+\xi)=X+\xi -d_\theta \alpha_{ij}(\xi) \in E_i|_{U_{ij}} .
\ena
Such $G_{ij}$ satisfies the cocycle condition
\bea
G_{ij}G_{jk}=G_{ik},
\ena
on $U_{ijk}$ due to (\ref{R-flux}).
Therefore, it defines the vector bundle over $M$ by 
\bea
E=\coprod_{x\in M} (TU_i)_0 \oplus (T^*U_i )_\theta /\sim. 
\ena
Since a $\beta$-gauge transformation preserves the anchor map, 
the inner product and the bracket (see (\ref{sym:beta-trf})), 
they all are globally well-defined on the quotient.
For example, the bracket on $U_i$ and $U_j$ are related by
\bea
[G_{ij}(X+\xi),G_{ij}(Y+\eta)]_i =G_{ij}([X+\xi, Y+\eta]_j ).
\ena
Therefore, the vector bundle $E$ is in fact a Courant algebroid.

It is apparent that $E$ satisfies the exact sequence (\ref{R-exact sequance}).
Here the map $\pi$ is induced by the projection 
$(TU_i)_0 \oplus (T^*U_i )_\theta \to (T^*U_i)_\theta$ to the second factor
and $\pi^*$ is the inclusion. 

As in the case of $H$-twist, the set of bivectors $\{\beta_i\}$ 
induces a bundle map $s: (T^*M)_\theta \to E$, locally defined by a $\beta$-transform as
\bea
s(\xi )=e^{\beta_i} (\xi) =\xi +\beta_i (\xi)
\ena
for $\xi \in T^*U_i$.
It follows form (\ref{R-flux}) 
that $s( (T^*M)_\theta)$ is globally well-defined as a vector bundle over $M$.
This map $s$ is in fact an isotropic splitting, since it satisfies 
$\pi\circ s (\xi)=\xi$ and $\langle s(\xi), s(\eta)\rangle =0$ for all $\xi, \eta \in (T^*M)_\theta$.
Therefore, $s$ induces the isotropic splitting $E= \pi^* ((TM)_0) \oplus s ((T^*M)_\theta)$ 
of $E$, 
and any section $A\in E$ can be uniquely expressed for $X\in TM$ and $\xi \in T^*M$ as
\bea
A=X+s(\xi).
\label{general elements}
\ena

\subsection{$R$-twisted bracket}
From this splitting, the structure of the Courant algebroid in 
$E=\pi^* ((TM)_0) \oplus s ((T^*M)_\theta)$ 
is translated to that in $ (TM)_0 \oplus (T^*M)_\theta$.
Since $s$ is a $\beta$-transformation $e^{\beta_i}$ locally, 
the anchor map and the inner product is unchanged from $ (TM)_0 \oplus (T^*M)_\theta$ 
(see (\ref{sym:beta-trf})):
\bea
&\rho (X+s(\xi))=\theta (\xi)=\rho (X+\xi), \quad
\langle X+s(\xi) ,Y+s(\eta) \rangle =\langle X+\xi ,Y+\eta \rangle.
\ena
The bracket on $\pi^* ((TM)_0) \oplus s ((T^*M)_\theta)$ 
is our bracket of sections of the form (\ref{general elements}).
We compute it locally as (see (\ref{sym:beta-trf}))
\bea
[X+s(\xi) ,Y+s(\eta)]
&=[e^{\beta_i} (X+\xi), e^{\beta_i} (Y+\eta)] \nn
&=e^{\beta_i} [X+\xi ,Y+\eta] +[\theta,\beta_i]_S (\xi,\eta)\nn
&=s([\xi,\eta]_\theta)+{\cal L}_\xi Y-{\cal L}_\eta X+\frac{1}{2}d_\theta (i_X\eta -i_Y\xi)
+(d_\theta \beta_i) (\xi,\eta).
\label{curvature calculation}
\ena
Hence, if we define the $R$-twisted bracket by
\bea
[X+\xi, Y+\eta]_R :=[X+\xi, Y+\eta] +R (\xi,\beta),
\ena
then we have 
\bea
[X+s(\xi) ,Y+s(\eta)]
&=(\pi^* \oplus s) ([X+\xi ,Y+\eta]_R ).
\ena
Therefore, as a Courant algebroid, 
$E=\pi^* ((TM)_0) \oplus s ((T^*M)_\theta)$
is equivalent to $(TM)_0 \oplus (T^*M)_\theta$ but with the $R$-twisted bracket.

\subsection{Poisson cohomology}

We can consider the cohomology based on $d_\theta$ (Poisson cohomology).
It controls the redundancy of $E$ constructed by $(R,\beta,\alpha)$,
just as in the presence of the $H$-flux. 
Here we discuss about this redundancy of the construction of the twisted Courant algebroid $E$.

For any other splitting $s':(T^*M)_\theta \to E$ of the same $E$,
the difference $s'-s$ should be an action of some global bivector $\beta \in \wedge^2 TM$,
in order to keep the transition function.
This means the splitting $s'$ is defined by the set of local bivectors of the form 
${\beta'}_i=\beta_i +\beta$, and they induce the flux $R'=R+d_\theta \beta$.
Thus, although the $R$-twisted bracket is changed,
it does not change the Poisson-cohomology class $[R']=[R]\in H_\theta^3 (M)$.
It is the analogue of the \v{S}evera class \cite{SeveraWeinstein}.
In fact, we see from (\ref{curvature calculation}) that
\bea
R(\xi,\eta,\zeta ) =2\langle [s(\xi ),s(\eta )], s(\zeta )\rangle,
\ena
and shown that the right hand side is a $d_\theta$-closed $3$-vector in general.
Therefore, similar to the case of $H$-fluxes, we conclude that 
$E$ is classified by the Poisson 3rd cohomology $[R] \in H^3_\theta (M)$.

Another redundancy comes form the choice of the data (\ref{R-flux}) keeping the same $R$.
Note that the set of bivectors $\{\beta_i\}$ in (\ref{R-flux}) 
gives the particular trivialization of the bundle $E$ with the splitting.
Keeping this structure and the bracket, the different choice of the data keeping the same $R$
is restricted to the local $\beta$-gauge transformation 
(\ref{local beta gauge symmetry}) defined on each open set $U_i$.
This changes the transition function to the equivalent one
\bea
&G'_{ij}(x) =\mat{1,d_\theta \Lambda_i (x),0,1}\mat{1,-d_\theta \alpha_{ij} (x),0,1}
\mat{1,-d_\theta \Lambda_j (x),0,1},
\ena
and thus changes $E$ to the isomorphic one.
Therefore, the construction of $E$ from the data (\ref{R-flux}) is in fact depends on $R$ up to a splitting-preserving isomorphism.

\subsection{As a quasi-Lie bialgebroid}

As we have seen, $s((T^*M)_\theta)$ is a maximally isotropic subbundle of $E$.
It is clear from (\ref{curvature calculation}) that 
the obstruction for it to be a Dirac structure is measured by the $R$-flux:
\bea
R(\xi,\eta)=[s(\xi), s(\eta)]- s([\xi,\eta]_\theta),
\ena
called the curvature of the splitting $s$.
Thus, $E=\pi^* ((TM)_0) \oplus s ((T^*M)_\theta)$ is not a Lie bialgebroid, but
this makes $E$ a {\it quasi-Lie bialgebroid}, a class of Courant algebroids.

In general, a quasi-Lie bialgebroid is defined by a triple $(A,\delta,\phi)$, 
consists of a Lie algebroid $A$, a degree $1$ derivation $\delta$ on the Gerstenhaber algebra
$(\Gamma(\wedge^\bullet A), \wedge,[\cdot,\cdot]_A)$ 
and a $A$-$3$-vector $\phi \in \Gamma(\wedge^3 A)$ 
such that $\delta^2=[\phi,\cdot]_A$ and $\delta \phi=0$.
See \cite{Roytenberg:2001am, LaurentGengoux:2007kh}
for more detail on quasi-Lie bialgebroids.

In our case, the Courant algebroid $E$ twisted by an $R$-flux defines a triple 
$(A,\, \delta,\, \phi)=((TM)_0, \,d_\theta, \,R)$.
Since the Lie algebroid $A=(TM)_0$ has the vanishing bracket, the corresponding 
Gerstenhaber bracket also vanishes $[\cdot,\cdot]_A=0$.
Thus the conditions on $\delta$ and $\phi$ are satisfied by $d_\theta^2 =0$ and $d_\theta R=0$.
Of course, it reduces to a Lie bialgebroid $((TM)_0, d_\theta, 0)=(TM)_0 \oplus (T^*M)_\theta$
for the vanishing $R$-flux.
Note that the $H$-twisted generalized tangent bundle is also a quasi-Lie bialgebroid 
of the form $(A,\delta,\phi)=(T^*M, d, H)$.\\

This completes the proposal of $R$-fluxes.
This is based on the $\beta$-gauge symmetry of the 
new Courant algebroid $(TM)_0 \oplus (T^*M)_\theta$,
and it is a complete analogy of $H$-fluxes.
Thus, an $R$-flux is geometric in the same sense that an $H$-fluxes is geometric.
We emphasize again that the validity of this proposal  will be justified in physics.
We end this section with a few remarks:

Any closed form admits a trivialization, since every manifold has a good cover and due to the Poincar\'e lemma.
Here we assume a particular trivialization or equivalently, the set of data $(R,\beta_i,\alpha_{ij})$. 
We do not know whether the Poisson version of the Poincar\'e lemma holds.
In other words, what is the notion of the ``good covering" in this case?

We here restrict ourselves to the data $(R,\beta_i,\alpha_{ij})$,
which corresponds to the data $(H,B_i,A_{ij})$ in the case of $H$-fluxes.
It is known that the latter can be promoted to the data of a $U(1)$-gerbe (with connection)
$(H,B_i,A_{ij}, n_{ijk})$ as a Deligne complex, where $n_{ijk} \in {\mathbb Z}$.
In this case, due to de Rham's theorem, $[H] \in H_{\rm dR}^3 (M)$ is an image of 
the \v Cech cohomology $\hat{H}^3 (M)$.
In our case, it is possible to consider the analogue of the 
$U(1)$-gerbe $(R,\beta_i,\alpha_{ij}, \nu_{ijk})$ formally,
but its relevance is not clear for us at present.

The situation becomes more symmetric  as $(M,\omega,H)\leftrightarrow (M,\theta,R)$, 
if we add a symplectic form $\omega$ in the left hand side.
This global closed $2$-form $\omega$ is absorbed into a shift $B_i \to B_i +\omega$ of local $B$-fields,
without changing its $H$-flux.
This is contrast to the different roles of $\theta$ and $\beta_i$.

The construction of $R$-fluxes in this section is straightforwardly 
extended to any Lie bialgebroid $A\oplus A^*$, 
with vanishing anchor and the Lie bracket for $A$.
It is a Courant algebroid, and a ``$\beta$-transformation" $e^\beta$ is a symmetry, 
where $\beta \in \wedge^2 A$ such that $d_{A^*} \beta=0$.
By the twisting using a global $A$-trivector ($A^*$-$3$-form) $R \in \wedge^3 A$, we obtain a 
twisted Courant algebroid.
$R$-fluxes and $H$-fluxes are particular cases of this general construction.

\section{Poisson-generalized geometry}

As we emphasized in the previous sections, the new Courant algebroid 
$(TM)_0 \oplus (T^*M)_\theta$ 
is a counterpart of the standard generalized tangent bundle $TM\oplus T^*M$,
where the role of $TM$ and $T^*M$ are completely interchanged.
Thus, we expect that all the mathematics concerning $TM\oplus T^*M$ has its counterpart in $(TM)_0\oplus (T^*M)_\theta$.
We call the latter as the Poisson-generalized geometry.
In this section, we briefly address the preliminary considerations about this geometry.
More detailed study is needed on each topic.

\subsection{Dirac structure}

A Dirac structure $L$ is defined in the same manner 
as in the standard generalized geometry.
That is, a Dirac structure $L\subset (TM)_0 \oplus (T^*M)_\theta$ 
is a maximally isotropic subbundle, and is involutive 
with respect to the new bracket $[L,L]\subset L$.
There are always two Dirac structures independent of the choice of a Poisson bivector $\theta$:
\begin{enumerate}
\item $L=(T^*M)_\theta$. Its bracket $[\cdot,\cdot]_\theta$ is a Lie bracket.
because of $\rho(L)=\theta(T^*M)$, the dimension of the leaf equals to the rank of $\theta$.
\item $L=(TM)_0$. Its bracket vanishes, and $\rho (L)=0$.
All leaves are $0$-dimensional.
\end{enumerate}
Contrary to the standard generalized geometry, even a simple subbundle such as $L={\rm span}\{\p_a,dx^i\}$ is not necessarily a Dirac structure, depending on the choice 
of the Poisson bivector.
Nevertheless, we can say some general statements analogous to those 
given in the standard generalized geometry:
\begin{enumerate}
\item
Let $\Delta \subset T^*M$ be a subbundle of $T^*M$ such that it is involutive 
$[\Delta, \Delta]_\theta \subset \Delta$ with respect to the Koszul bracket.
Then, $L=\Delta \oplus {\rm Ann}(\Delta) $ is a Dirac structure of $(TM)_0\oplus (T^*M)_\theta$.\\
{\it Proof.}
$L$ is apparently maximally isotropic. 
The involutivity condition $[L,L]\subset L$ reduces to
$[\Delta, {\rm Ann}(\Delta)] \subset {\rm Ann}(\Delta)$, because
$[\Delta, \Delta]_\theta \subset \Delta$ and $[{\rm Ann}(\Delta), {\rm Ann}(\Delta)] =0$.
This condition means for arbitrary $\xi , \eta\in \Delta$ and $X\in  {\rm Ann}(\Delta) $ 
\bea
0=\langle [\xi,X],\eta \rangle ,
\ena
but it is rewritten as
\bea
0=\langle [\xi,\eta]_\theta,X \rangle,
\ena
which is automatically satisfied by definition.
{\it (End of the proof)}
\item
Given a Dirac structure $L$, its deformation $L_{\cal F}=e^{{\cal F}}L$ 
by a $L$-$2$ form ${\cal F}\in \wedge^2 L^*$ is still a Dirac structure iff
the Maurer-Cartan type equation
$d_L {\cal F} +\frac{1}{2}[{\cal F},{\cal F}]_{L^*}=0$ 
is satisfied \cite{LWX}.
For $L=(T^*M)_\theta$, ${\cal F}$ is a bivector such that $d_\theta {\cal F}=0$.
For $L=(TM)_0$, ${\cal F}$ is a $2$-form such that $[{\cal F},{\cal F}]_\theta=0$
($[\cdot,\cdot]_\theta$ is the Gerstenhaber bracket extending the Koszul bracket.).
\end{enumerate}

In the papers \cite{AMW,ASW,Koerber}, Dirac structures are identified 
with D-branes (with fluctuations).
It is interesting to investigate the Dirac structures here in this context.

\subsection{Generalized Riemannian structure}

As in the standard Courant algebroid, we define a generalized Riemannian structure as 
a maximal-positive definite subbundle $C_+ \subset (TM)_0 \oplus (T^*M)_\theta$.
Since this definition depends only on $TM\oplus T^*M$ as a vector bundle, and the bilinear form 
is independent of the bracket, 
a generalized Riemannian structure $C_+$ of the standard tangent bundle
becomes automatically that of the new Courant algebroid.
In other words, two Courant algebroids share the same $C_+$.
Therefore, $C_+$ is given by a graph of a map $g+B:TM \to T^*M$,
\bea
C_+ =\{ X +(g+B) (X)~|~X\in TM\}.
\label{metric from TM}
\ena
As is emphasized in our previous papers \cite{ASW, AMW}, however, 
there are various ways to represent $C_+$ as graphs.
In particular, $C_+$ can be seen from $T^*M$ as
\bea
C_+ =\{ \xi +(G+\beta) (\xi)~|~\xi\in T^*M\},
\label{metric from T*M}
\ena
where $G+\beta =(g+B)^{-1} : T^*M \to TM$ is the inverse map.
Two representations (\ref{metric from TM}) and (\ref{metric from T*M}) of $C_+$ 
are equivalent if the fluxes are absent.

However, in the presence of the fluxes, the situation is changed. 
In the presence of an $H$-flux, the representation (\ref{metric from TM}) is natural, 
since an $H$-twisting requires to replace $B$ with a local $2$-form $B_i$ while
it does not affect the symmetric part $g$.
In other words, a Riemannian manifold $(M,g)$ is unchanged 
regardless of the presence of $H$-fluxes.
This compatibility of the generalized metric with $H$-twisting is emphasized in \cite{Hitchin2}.
On the other hand, in the different representation (\ref{metric from T*M}), an $H$-twisting affects 
both the symmetric part $G$ and the skew-symmetric part $\beta$, 
so that $G$ is non-trivially glued by local $B$-gauge transformations.

Similarly, in the presence of a $R$-flux, the representation (\ref{metric from T*M}) is natural,
since now a $R$-twisting affects only $\beta$, kept fixed the symmetric part $G$.
Here $G$ is a fiber metric on $T^*M$ defining a Riemannian manifold.

\subsection{Clifford module and pure spinor}

For any section $X+\xi \in TM\oplus T^*M$ of the vector bundle $TM\oplus T^*M$, 
its Clifford action either on differential forms $\wedge^\bullet T^*M$ 
or on polyvectors $\wedge^\bullet TM$ is defined by
\bea
&\gamma_{X+\xi} \omega =i_X \omega +\xi\wedge \omega, ~~\omega \in \wedge^\bullet T^*M, \nn
&\gamma_{X+\xi} u =X \wedge V +i_\xi V, ~~V \in \wedge^\bullet TM, 
\ena
which satisfies
\bea
\{\gamma_{X+\xi}, \gamma_{Y+\eta}\} =2\langle X+\xi ,Y+\eta \rangle.
\ena
In this sense, both differential forms $\wedge^\bullet T^*M$ 
and polyvectors $\wedge^\bullet TM$ can be identified as spinors.

In the generalized geometry with the Courant bracket on $TM\oplus T^*M$,
differential forms $\wedge^\bullet T^*M$ are considered to be spinors.
This is because the analogue of the Cartan relation among $\gamma_A$, $d_H$ 
and ${\cal L}_A$ holds:
\bea
&\{ \gamma_A, \gamma_B\}=2\langle A,B\rangle, \quad
\{ d_H ,\gamma_A\}={\cal L}_A, \quad
[{\cal L}_A, \gamma_B ]=\gamma_{A\circ_H B},\nn
&[{\cal L}_A, {\cal L}_B ]={\cal L}_{[A,B]_H},\quad
[d_H,{\cal L}_A ]=0,
\ena
where $A,B \in TM\oplus T^*M$, $d_H=d+H\wedge$, and
${\cal L}_{X+\xi}\omega = {\cal L}_X \omega +(d\xi +i_X H) \wedge \omega $.
$A\circ_H B$ is the $H$-twisted Dorfman bracket.
We refer to these equations as the Clifford-Cartan relation.
For a given spinor $\varphi \in \wedge^\bullet T^*M$, nonvanishing everywhere, 
its annihilator bundle $L_\varphi =\{X+\xi \in TM\oplus T^*M|\gamma_{X+\xi}\varphi =0\}$
is defined.
Then, $\varphi $ is called a pure spinor if $L_\varphi $ is a maximally isotropic subbundle
of $TM\oplus T^*M$.
Moreover, $L_\varphi$ is involutive, if $d_H \varphi =0$. 
Therefore, there is a correspondence between a Dirac structure $L_\varphi $ and a pure spinor $\varphi $.

Now let us turn the discussion on our Courant algebroid.
Since the roles of $TM$ and $T^*M$ are exchanged, it is natural to regard 
polyvectors $\wedge^\bullet TM$ as spinors.
In fact, the Clifford-Cartan relation among $\gamma_A$, $d_R$ and ${\cal L}_A$ holds:
\bea
&\{ \gamma_A, \gamma_B\}=2\langle A,B\rangle, \quad
\{ d_R ,\gamma_A\}={\cal L}_A, \quad
[{\cal L}_A, \gamma_B ]=\gamma_{A\circ_R B},\nn
&[{\cal L}_A, {\cal L}_B ]={\cal L}_{[A,B]_R},\quad
[d_R,{\cal L}_A ]=0,
\label{A Clifford-Cartan}
\ena
where 
\bea 
&d_R=d_\theta +R\wedge, \quad 
{\cal L}_{X+\xi}V = {\cal L}_\xi V+(d_\theta X +i_\xi R) \wedge V ,\nn
&(X+\xi)\circ_R (Y+\eta) = [\xi, \eta]_\theta +{\cal L}_{\xi} Y - i_\eta d_\theta X -i_\eta i_\xi R
\ena
The proof is given in the appendix \ref{app fin}.
Since the algebraic relation here is the same as that of the Courant bracket,
the same argument holds concerning pure spinors.
Namely, a pure spinor $\varphi \in \wedge^\bullet TM$ is a polyvector and if $d_R \varphi =0$ then
its annihilator bundle $L_\varphi $ is a Dirac structure.
The difference is that we should work with $d_\theta$ and thus with the Poisson cohomology.

\section{Conclusion and Discussion}

In this paper, we studied the new Courant algebroid $(TM)_0\oplus (T^*M)_\theta$ 
defined on Poisson manifolds, 
as an analogue of the generalized tangent bundle $TM\oplus T^*M$ in the generalized geometry.
It is an extension of the Lie algebroid of a Poisson manifold $(T^*M)_\theta$, and the symmetry 
consists of $\beta$-diffeomorphisms and $\beta$-transformations.
We then proposed a definition of $R$-fluxes 
as a twist of the new Courant algebroid, having the analogous structure with the $H$-flux.
It is a $d_\theta$-closed global $3$-vector and used to twist the Courant algebroid 
$(TM)_0\oplus (T^*M)_\theta$.
It is an abelian field strength of local bivector gauge potentials $\{\beta_i\}$.
We also briefly discussed about the Poisson-generalized geometry based on 
$(TM)_0\oplus (T^*M)_\theta$, such as Dirac structures, generalized Riemannian structures and
pure spinors.

In our construction, the $R$-flux is completely geometric 
but a space with an $R$-flux is usually considered as a non-geometric space in the literature.
We do not know the reason of this discrepancy at present, and it should be investigated further.
Note that we have used an unusual tangent bundle $(TM)_0$ with vanishing Lie bracket.
A non-geometric nature may arise when $(TM)_0$ is treated as an ordinary tangent bundle 
$TM$ with non-vanishing Lie bracket.

Since we focused on the definition of $R$-fluxes in this paper,
there are many related topics and unsolved questions. 
Along the approach of this paper, 
we would like to define another non-geometric flux, a $Q$-flux  \cite{AMW2}.
It will be important to understand the T-duality chain in fully geometric way.
It also needs more detailed study on the Poisson-generalized geometry,
such as generalized complex structures.

As emphasized, our proposal is mainly based on 
the structure of Courant algebroids.
Thus the most important question is whether our 
$R$-fluxes are realized in string theory or supergravity.
In the case of $H$-fluxes, $H$ should be quantized, 
since it appears in the WZW-term in the string worldsheet theory.
Similarly, $R$ should also be quantized when it is realized as a background flux 
in the string worldsheet theory 
or membrane worldvolume theory \cite{Halmagyi:2008dr, Mylonas:2012pg}.
It is interesting to see whether our $R$-fluxes are consistent with these formulations.
There, the $U(1)$-gerbe analogue of $R$-fluxes would play the role.

It is also possible to consider a gravity theory in our Courant algebroid with $R$-fluxes.
The formulation should be based on the differential geometry of $(T^*M)_\theta$.
In this sense, it would touch upon the work \cite{Blumenhagen:2012nt, Andriot:2013xca}, where 
the gravity theory based on $\beta$-diffeomorphisms is constructed and it is physically equivalent 
to the original supergravity.
However, as emphasized above, since an $R$-twisting is 
very different form an $H$-twisting in nature,
the resulting theory is in general not expected to be equivalent to the ordinary supergravity,
except for special cases where two twistings are related.


\section*{Acknowledgments}
Authors would like to thank 
the members of the particle theory and cosmology group, 
in particular U.~Carow-Watamura for helpful comments and discussions.
S.~W. would also like to thank N.~Ikeda and Y.~Maeda for valuable discussions.
H.~M. is supported by Tohoku University
Institute for International Advanced Research and Education.

\appendix

\section{Proof of the third equation of (\ref{gege beta-diffeo})  \label{app beta Lie}}
We will prove the third equation of (\ref{gege beta-diffeo}),
\bea
{\cal L}_{\zeta} X ={\cal L}_{\theta(\zeta)}X +\theta (i_X d\zeta),
\ena
in the the components calculation.
Because of
\bea
[\theta,X]_S 
&=\left[\textstyle{\frac{1}{2}}\theta^{\mu\nu} \p_\mu\wedge\p_\nu, X^\alpha \p_\alpha \right]_S \nn
&=\textstyle{\frac{1}{2}}\left[\theta^{\mu\nu} \p_\mu, X^\alpha \p_\alpha \right]_S \wedge\p_\nu
-\textstyle{\frac{1}{2}}\left[\p_\nu, X^\alpha \p_\alpha \right]_S \wedge \theta^{\mu\nu} \p_\mu \nn
&=\textstyle{\frac{1}{2}} \theta^{\mu\nu} \p_\mu X^\alpha \p_\alpha \wedge \p_\nu 
-\textstyle{\frac{1}{2}}\theta^{\mu\nu} \p_\nu X^\alpha \p_\alpha \wedge \p_\mu 
-\textstyle{\frac{1}{2}} X^\alpha\p_\alpha \theta^{\mu\nu} \p_\mu \wedge \p_\nu \nn
&=\theta^{\mu\nu} \p_\mu X^\alpha \p_\alpha \wedge \p_\nu 
-\textstyle{\frac{1}{2}} X^\alpha\p_\alpha \theta^{\mu\nu} \p_\mu \wedge \p_\nu,
\ena
so we have
\bea
i_\zeta d_\theta X 
&=i_\zeta [\theta,X]_S \nn
&=\theta^{\mu\nu} \p_\mu X^\alpha \zeta_\alpha \p_\nu 
-\theta^{\mu\nu} \p_\mu X^\alpha \zeta_\nu \p_\alpha 
-X^\alpha\p_\alpha \theta^{\mu\nu} \zeta_\mu \p_\nu \nn
&=\left(\theta^{\mu\rho} \p_\mu X^\alpha \zeta_\alpha 
+\theta^{\mu\nu} \p_\nu X^\rho \zeta_\mu 
-X^\alpha\p_\alpha \theta^{\mu\rho} \zeta_\mu
\right) \p_\rho.
\ena
Next, we compute
\bea
d_\theta i_\zeta  X 
&=-\theta (d (i_\zeta X)) \nn 
&=-\theta^{\mu\nu} \p_\mu (\zeta_\alpha X^\alpha) \p_\nu \nn
&=-\theta^{\mu\rho} (\p_\mu \zeta_\alpha X^\alpha +\zeta_\alpha \p_\mu X^\alpha) \p_\rho.
\ena
Therefore, the l.h.s. is written as
\bea
{\cal L}_{\zeta} X 
&=i_\zeta d_\theta X +d_\theta i_\zeta  X \nn
&=\left(\theta^{\mu\rho} \p_\mu X^\alpha \zeta_\alpha 
+\theta^{\mu\nu} \p_\nu X^\rho \zeta_\mu 
-X^\alpha\p_\alpha \theta^{\mu\rho} \zeta_\mu
\right) \p_\rho
-\theta^{\mu\rho} (\p_\mu \zeta_\alpha X^\alpha +\zeta_\alpha \p_\mu X^\alpha) \p_\rho \nn
&=\left(
\theta^{\mu\nu} \p_\nu X^\rho \zeta_\mu 
-\theta^{\mu\rho} \p_\mu \zeta_\alpha X^\alpha 
-X^\alpha\p_\alpha \theta^{\mu\rho} \zeta_\mu
\right) \p_\rho.
\ena
On the other hand, the r.h.s. is computed as follows.
The first term is written as
\bea
{\cal L}_{\theta(\zeta)}X
&=[\theta(\zeta), X]_S 
=[\theta^{\mu\nu}\zeta_\mu \p_\nu, X^\alpha\p_\alpha]_S \nn
&=\theta^{\mu\nu}\zeta_\mu \p_\nu X^\alpha\p_\alpha 
-X^\alpha\p_\alpha (\theta^{\mu\nu}\zeta_\mu) \p_\nu \nn
&=\left(
\theta^{\mu\nu} \zeta_\mu \p_\nu X^\rho
-X^\alpha\p_\alpha \theta^{\mu\rho} \zeta_\mu
-X^\alpha \theta^{\mu\rho} \p_\alpha \zeta_\mu
\right) \p_\rho,
\ena
and the second term is 
\bea
\theta (i_X d\zeta)
&=\theta \left(i_X \left(\textstyle{\frac{1}{2}}(\p_\mu\zeta_\nu-\p_\nu \zeta_\mu ) 
dx^\mu\wedge dx^\nu\right)\right) \nn
&=\theta \left( X^\mu (\p_\mu\zeta_\nu-\p_\nu \zeta_\mu ) dx^\nu\right)
=\theta \left( X^\nu (\p_\nu\zeta_\mu-\p_\mu \zeta_\nu ) dx^\mu\right)\nn
&=\theta^{\mu\rho} X^\nu (\p_\nu\zeta_\mu-\p_\mu \zeta_\nu ) \p_\rho.
\ena
Summing up, we obtain
\bea
{\cal L}_{\theta(\zeta)}X+\theta (i_X d\zeta)
&=\left(
\theta^{\mu\nu} \zeta_\mu \p_\nu X^\rho
-X^\alpha\p_\alpha \theta^{\mu\rho} \zeta_\mu
-X^\alpha \theta^{\mu\rho} \p_\alpha \zeta_\mu
+\theta^{\mu\rho} X^\nu (\p_\nu\zeta_\mu-\p_\mu \zeta_\nu )
\right) \p_\rho\nn
&=\left(
\theta^{\mu\nu} \zeta_\mu \p_\nu X^\rho
-X^\alpha\p_\alpha \theta^{\mu\rho} \zeta_\mu
-\theta^{\mu\rho} X^\nu \p_\mu \zeta_\nu
\right) \p_\rho\nn
&=\left(
\theta^{\mu\nu} \p_\nu X^\rho \zeta_\mu 
-\theta^{\mu\rho} \p_\mu \zeta_\alpha X^\alpha 
-X^\alpha\p_\alpha \theta^{\mu\rho} \zeta_\mu
\right) \p_\rho.
\ena
Thus, the equation is proved.

\section{Proofs of ($\ref{sym:beta-diffeo}$) \label{app beta diffeo}}

The proof of the first equation is shown as follows.
\bea
\langle {\cal L}_{\zeta} (X+\xi), Y+\eta \rangle +\langle X+\xi, {\cal L}_{\zeta}(Y+\eta)\rangle 
=& \textstyle{\frac{1}{2}}(i_\eta {\cal L}_{\zeta} X+i_{{\cal L}_{\zeta} \xi} Y
+ i_\xi {\cal L}_{\zeta} Y+i_{{\cal L}_{\zeta} \eta}X) \nn
=&\textstyle{\frac{1}{2}}{\cal L}_{\zeta} (i_\eta  X +i_\xi Y) \nn
=&{\cal L}_{\zeta}\langle X+\xi, Y+\eta \rangle,
\ena
where we used $i_{{\cal L}_{\zeta} \xi}=i_{[\zeta, \xi]_\theta}=[{\cal L}_{\zeta}, i_\xi]$ 
given in (\ref{A Cartan}).
Next, let us prove the third equation.
The r.h.s. is 
\bea
{\cal L}_{\zeta}[X+\xi,Y+\eta ]
&={\cal L}_{\zeta}[\xi,\eta ]_\theta +{\cal L}_{\zeta}{\cal L}_{\xi}Y -{\cal L}_{\zeta}{\cal L}_{\eta}X
+\textstyle{\frac{1}{2}}{\cal L}_{\zeta}d_\theta (i_X\eta -i_Y\xi ).
\ena
By using the relations following from (\ref{A Cartan}),
\bea
&{\cal L}_{\zeta}[\xi,\eta ]_\theta =[\zeta,[\xi,\eta ]_\theta]_\theta
=[{\cal L}_{\zeta}\xi,\eta ]_\theta + [\xi,{\cal L}_{\zeta}\eta ]_\theta,\nn
&{\cal L}_{\zeta}{\cal L}_{\xi}Y
={\cal L}_{\xi}{\cal L}_{\zeta}Y +{\cal L}_{[\zeta,\xi]_\theta}Y,
\ena
the r.h.s. is further rewritten as
\bea
{\cal L}_{\zeta}[X+\xi,Y+\eta ]
&=[{\cal L}_{\zeta}\xi,\eta ]_\theta + [\xi,{\cal L}_{\zeta}\eta ]_\theta
+{\cal L}_{\xi}{\cal L}_{\zeta}Y +{\cal L}_{[\zeta,\xi]_\theta}Y
-{\cal L}_{\eta}{\cal L}_{\zeta}X -{\cal L}_{[\zeta,\eta]_\theta}X \nn
&+\textstyle{\frac{1}{2}}d_\theta i_{\zeta}d_\theta (i_X\eta -i_Y\xi ). \label{blhs}
\ena
On the other hand, the first term in the l.h.s. is
\bea
[{\cal L}_{\zeta}(X+\xi),Y+\eta ]
&=[{\cal L}_{\zeta}\xi,\eta ]_\theta + {\cal L}_{[\zeta,\xi]_\theta}Y-{\cal L}_{\eta}({\cal L}_{\zeta}X)
+\textstyle{\frac{1}{2}}d_\theta  (i_{{\cal L}_{\zeta}X}\eta -i_Y({\cal L}_{\zeta}\xi )),
\ena
and similar for the second term.
Thus, the l.h.s gives
\bea
&[{\cal L}_{\zeta}(X+\xi),Y+\eta ]+[X+\xi,{\cal L}_{\zeta}(Y+\eta) ]\nn
=&[{\cal L}_{\zeta}\xi,\eta ]_\theta + {\cal L}_{[\zeta,\xi]_\theta}Y-{\cal L}_{\eta}({\cal L}_{\zeta}X)
+\textstyle{\frac{1}{2}}d_\theta  (i_{{\cal L}_{\zeta}X}\eta -i_Y({\cal L}_{\zeta}\xi )) \nn
+&[\xi,{\cal L}_{\zeta}\eta ]_\theta +{\cal L}_{\xi}({\cal L}_{\zeta}Y) - {\cal L}_{[\zeta,\eta]_\theta}X
+\textstyle{\frac{1}{2}}d_\theta  (i_{X}({\cal L}_{\zeta}\eta ) -i_{{\cal L}_{\zeta}Y} \xi ). \label{brhs}
\ena
Then except for the $d_\theta$-exact terms, 
it is apparent that (\ref{blhs}) and (\ref{brhs}) coincide. 
Moreover, the $d_\theta$-exact terms are also the same, since
\bea
i_{{\cal L}_{\zeta}X}\eta +i_{X}({\cal L}_{\zeta}\eta )
={\cal L}_{\zeta}(i_X\eta )
=i_\zeta d_\theta (i_X\eta ).
\ena
Here we used the formula of the action of the
Lie derivative on a function, ${\cal L}_{\zeta}f=i_\zeta d_\theta f$.

Finally, we check the second equation.
The l.h.s is given as
$
\rho ({\cal L}_{\zeta} (X+\xi)) =\theta ({\cal L}_{\zeta} \xi ),
$
while the r.h.s is 
$
{\cal L}_{\zeta} (\rho (X+\xi)) =({\cal L}_{\zeta}\theta )(\xi) +\theta ({\cal L}_{\zeta} \xi ),
$
so that the equation is satisfied if
\bea
{\cal L}_{\zeta}\theta =d_\theta i_\zeta \theta =d_\theta \theta (\zeta) =0.
\ena

\section{Proof of the third equation of (\ref{sym:beta-trf}) \label{app beta transf}}

To this end we will show that
\bea
&e^{\beta}[X+\xi,Y+\eta ]=[e^{\beta}(X+\xi),e^{\beta}(Y+\eta) ]+[\theta, \beta]_S (\xi,\eta)
\label{second_action}
\ena
then, a $\beta$-transformation is a symmetry if $d_\theta \beta =[\theta,\beta]_S=0$.
The l.h.s. is written as
\bea
e^{\beta}[X+\xi,Y+\eta ]
&=[X+\xi,Y+\eta ]+\beta ([\xi,\eta]_\theta).
\ena
while the r.h.s. is
\bea
[e^{\beta}(X+\xi),e^{\beta}(Y+\eta) ]
&=[X+\xi +\beta(\xi),Y+\eta +\beta(\eta) ]\nn
&=[X+\xi,Y+\eta ] +{\cal L}_{\xi}\beta(\eta) -{\cal L}_{\eta}\beta(\xi) 
+\textstyle{\frac{1}{2}}d_\theta (i_{\beta(\xi)} \eta -i_{\beta(\eta)}\xi ).
\ena
By using the formula ${\cal L}_{\zeta} X ={\cal L}_{\theta(\zeta)}X +\theta (i_X d\zeta)$,
we have
\bea
{\cal L}_{\xi}\beta(\eta) 
&={\cal L}_{\theta(\xi)}\beta(\eta)+\theta (i_{\beta(\eta)} d\xi)\nn
&=[\theta(\xi),\beta(\eta)]_\theta +\theta (i_{\beta(\eta)} d\xi),
\ena
By using $d_\theta f=-\theta(df)$, we have
\bea
d_\theta i_{\beta(\xi)} \eta  =d_\theta (\beta(\xi, \eta))
=-\theta (d(\beta(\xi, \eta)))
\ena
Substituting these, the r.h.s. becomes
\bea
&[e^{\beta}(X+\xi),e^{\beta}(Y+\eta) ] \nn
=&[X+\xi,Y+\eta ] 
+[\theta(\xi),\beta(\eta)]_\theta -[\theta(\eta),\beta(\xi)]_\theta
+\theta (i_{\beta(\eta)} d\xi -i_{\beta(\xi)} d\eta - d(\beta (\xi, \eta )))\nn
=&[X+\xi,Y+\eta ] 
+[\theta(\xi),\beta(\eta)]_\theta +[\beta(\xi),\theta(\eta)]_\theta
- \theta ([\xi,\eta]_\beta ),
\ena
where in the last line we define $[\xi,\eta]_\beta$ by the same formula as the Koszul bracket
for an arbitrary bivector $\beta$ (It is not a Lie bracket but we do not use this property.).
Then, by using 
\bea
[(\theta+\beta)(\xi),(\theta+\beta)(\eta)]_S 
=[\theta(\xi),\theta(\eta)]_S+[\theta(\xi),\beta(\eta)]_S
+[\beta(\xi),\theta(\eta)]_S+[\beta(\xi),\beta(\eta)]_S,
\ena
it is further rewritten as
\bea
&[e^{\beta}(X+\xi),e^{\beta}(Y+\eta) ] \nn
=&[X+\xi,Y+\eta ] 
+[(\theta+\beta)(\xi),(\theta+\beta)(\eta)]_S 
-[\theta(\xi),\theta(\eta)]_S -[\beta(\xi),\beta(\eta)]_S
- \theta ([\xi,\eta]_\beta ),
\ena
To rewrite it further, we use a formula
\bea
[\beta(\xi),\beta(\eta)]_S =\beta([\xi,\eta]_\beta )+\frac{1}{2}[\beta,\beta]_S (\xi,\eta)
\ena
which is valid for any bivector $\beta$. In particular,
\bea
[(\theta+\beta)(\xi),(\theta+\beta)(\eta)]_S 
&=(\theta+\beta) ([\xi,\eta]_{\theta+\beta}) +\frac{1}{2}[\theta+\beta,\theta+\beta]_S (\xi,\eta)\nn
&=(\theta+\beta) ([\xi,\eta]_\theta+[\xi,\eta]_\beta) 
+\frac{1}{2}[\theta+\beta,\theta+\beta]_S (\xi,\eta).
\ena
Then, we finally obtain
\bea
&[e^{\beta}(X+\xi),e^{\beta}(Y+\eta) ] \nn
=&[X+\xi,Y+\eta ] 
+(\theta+\beta) ([\xi,\eta]_\theta+[\xi,\eta]_\beta) 
-\theta ([\xi,\eta]_\theta) -\beta ([\xi,\eta]_\beta) - \theta ([\xi,\eta]_\beta ) \nn
&+\frac{1}{2}[\theta+\beta,\theta+\beta]_S (\xi,\eta)
-\frac{1}{2}[\theta,\theta]_S (\xi,\eta) -\frac{1}{2}[\beta,\beta]_S (\xi,\eta)\nn
=&[X+\xi,Y+\eta ] 
+\beta ([\xi,\eta]_\theta)+ [\theta,\beta]_S (\xi,\eta).
\ena

\section{Review on twisting of $TM\oplus T^*M$ with $H$-flux \label{app H-flux}}

When there is a $H$-flux, one can define the 
corresponding Courant algebroid $(E,\rho,[\cdot,\cdot])$ 
from $TM\oplus T^*M$ by twist as follows 
\cite{Gualtieri, SeveraWeinstein, Rogers}:
\begin{enumerate}

\item[1)] Take a good cover $\{U_i \}$ of $M$.
Before twisting, a global section of $TM\oplus T^*M$ satisfies 
$X_i +\xi_i =X_j +\xi_j $
on a overlap $U_{ij}=U_i \cap  U_j$.

\item[2)] Modify the gluing condition to 
$X_i +\xi_i =X_j +\xi_j -dA_{i j }(X_j )$
for a set of 1-forms $A_{i j } \in T^* U_{ij }$.
Note that $T^*M$ is twisted by local $B$-gauge transformations.

\item[3)] Define a bundle $E=\amalg_{i} \left(TU_i \oplus T^*U_i\right) /\sim$ by a 
standard clutching construction.
Then, $(E, \rho, [\cdot,\cdot])$ is a Courant algebroid, because the $B$-gauge transformation preserves both the anchor 
$\rho$ and the bracket $[\cdot,\cdot]$.

\end{enumerate}
This twisting defines an exact Courant algebroid
\begin{align}
0\to T^*M \xrightarrow{\rho^*} E \xrightarrow{\rho} TM \to 0. \label{twisting}
\end{align}
with an isotropic splitting $s: TM \to E$.
That is $E= s(TM)\otimes \rho^* (T^*M)$.
Locally, the splitting is given by a local $B$-transform as
\bea
s_i (X)=e^{B_i} (X) =X+B_i (X),
\ena
where $B_i \in \wedge^2 T^* U_i$.
In order that it is globally defined, it should satisfy $s_i (X)= s_j (X)$ on $U_{ij}$.
Taking into account the gluing condition 2), it leads to conditions 
$B_j = B_i + dA_{ij}$ for local $2$-forms.
It also implies that $H:=dB_i$ on $M$ is a global closed $3$-form.

Thus, we need a data $(H, B_i, A_{ij})$ to construct $E$.
More specifically, 
it is known that the geometric object corresponding to a closed 3-form $H$ flux is a $U(1)$-gerbe with connection,
when its cohomology class $[H]$ is in the integer cohomology $H^3(M;{\mathbb Z})$.
It is defined by a set $(H, B_i, A_{ij}, \Lambda_{ijk})$
in the \v{C}ech-de Rham double complex,
with a set of equations
\begin{align}
U_i:~~ &H = dB_i , \nn
U_{ij}:~~ &B_j - B_i = dA_{ij} ,\nn
U_{ijk}:~~ &A_{ij} + A_{jk} + A_{ki} = d \Lambda_{ijk} ,\nn
U_{ijkl}:~~ &\Lambda_{jkl} - \Lambda_{ikl} + \Lambda_{ijl} - \Lambda_{ijk} = n_{ijkl}.
\end{align}

This $H$-twisting is also regarded as a change of the Courant bracket of $TM\oplus T^*M$ to 
the $H$-twisted Courant bracket.
To see this recall that the relation
\bea
[e^{B_i} (X+\xi) , e^{B_i} (Y+\eta) ]
=e^{B_i} [X+\xi, Y+\eta ] + i_X i_Y d{B_i}
\label{local B-trf identity}
\ena
is still true for local $B$-transformations.
Therefore, if we define an $H$-twisted Courant bracket
\begin{align}
[X+\xi, Y+\eta ]_H=[X+\xi, Y+\eta ]+i_X i_Y H,
\label{H Courant}
\end{align}
then we have locally
\begin{align}
e^{B_i} [X+\xi, Y+\eta ]_H =[e^{B_i} (X+\xi) , e^{B_i} (Y+\eta) ],
\end{align}
and globally
\bea
[X+s(\xi) ,Y+s(\eta)]
&=(\rho^* \oplus s) ([X+\xi ,Y+\eta]_H ).
\ena
This defines an isomorphism of Courant algebroids
\begin{align}
(TM\oplus T^*M, \rho, [\cdot,\cdot]_H) \xrightarrow{} 
(E, \rho, [\cdot,\cdot]).
\label{local B iso}
\end{align}

We end this section with a remark about global $B$-transformations.
The another choice of the splitting  $s^\prime$ should differ from $s$ by a
$B$-transformation with a global $2$-form $b$
and change $E=s'(TM)\otimes \rho^*(T^*M)$, where $s'_i (X)=X+(B_i +b)(X)$.
It leads to the twisted bracket $[\cdot,\cdot]_{H+db}$ but does not change the cohomology class
in $H^3_{\rm dR}(M)$.

\section{Proof of (\ref{A Clifford-Cartan}) \label{app fin}}
The second equation is obtained as
\bea
\{ d_R ,\gamma_{X+\xi}\} V
&=d_R (X\wedge V +i_\xi V) +\gamma_{X+\xi} (d_\theta V +R\wedge V)\nn
&=d_\theta  (X\wedge V +i_\xi V) +R\wedge  (X\wedge V +i_\xi V)
+X\wedge (d_\theta V +R\wedge V) +i_\xi (d_\theta V +R\wedge V)\nn
&=(d_\theta  X)\wedge V +d_\theta (i_\xi V) +i_\xi (d_\theta V) +(i_\xi R)\wedge V \nn
&={\cal L}_\xi V +(d_\theta X +i_\xi R) \wedge V \nn
&={\cal L}_{X+\xi}V.
\ena
This is indeed the definition of ${\cal L}_{X+\xi}$.
The last (fifth) equation is shown as
\bea
[d_R,{\cal L}_{X+\xi} ]V
&=(d_\theta +R\wedge ) ({\cal L}_\xi V +(d_\theta X) \wedge V +(i_\xi R) \wedge V ) \nn
&-({\cal L}_\xi +(d_\theta X)\wedge   +(i_\xi R) \wedge ) (d_\theta V+R\wedge V) \nn
&=[d_\theta, {\cal L}_\xi ]V +(d_\theta i_\xi R) \wedge V -({\cal L}_\xi R) \wedge V \nn
&=0,
\ena
where $[d_\theta, {\cal L}_\xi ]=0$ and ${\cal L}_\xi R=d_\theta i_\xi R$ are used.
The third equation is shown as
\bea
[{\cal L}_{X+\xi}, \gamma_{Y+\eta} ]V
&={\cal L}_{X+\xi} (Y\wedge V+ i_\eta V) 
-\gamma_{Y+\eta} ({\cal L}_\xi V +(d_\theta X+i_\xi R) \wedge V)\nn
&={\cal L}_{\xi} (Y\wedge V+ i_\eta V)  +(d_\theta X+i_\xi R) \wedge (Y\wedge V+ i_\eta V)\nn
&-Y\wedge ({\cal L}_\xi V +(d_\theta X+i_\xi R) \wedge V) 
-i_\eta ({\cal L}_\xi V +(d_\theta X+i_\xi R) \wedge V) \nn
&=({\cal L}_{\xi} Y)\wedge V +[{\cal L}_{\xi}, i_\eta ]V -(i_\eta (d_\theta X +i_\xi R)) \wedge V \nn
&=i_{[\xi, \eta]_\theta} V +({\cal L}_{\xi} Y - i_\eta d_\theta X -i_\eta i_\xi R) \wedge V \nn
&=\gamma_{(X+\xi)\circ_R (Y+\eta) }V,
\ena
where $[{\cal L}_{\xi}, i_\eta ]=i_{[\xi, \eta]_\theta}$ and
\bea
(X+\xi)\circ_R (Y+\eta) &= [\xi, \eta]_\theta +{\cal L}_{\xi} Y - i_\eta d_\theta X -i_\eta i_\xi R
\ena
are used.
The fourth equation is shown as
\bea
[{\cal L}_{X+\xi}, {\cal L}_{Y+\eta} ]V
&={\cal L}_{X+\xi} ({\cal L}_\eta V+(d_\theta Y +i_\eta R) \wedge V) 
-{\cal L}_{Y+\eta}({\cal L}_\xi V+(d_\theta X +i_\xi R) \wedge V) \nn
&={\cal L}_{\xi} ({\cal L}_\eta V+(d_\theta Y +i_\eta R) \wedge V) 
+(d_\theta X +i_\xi R) \wedge ({\cal L}_\eta V+(d_\theta Y +i_\eta R) \wedge V) \nn
&-{\cal L}_{\eta}({\cal L}_\xi V+(d_\theta X +i_\xi R) \wedge V) 
-(d_\theta Y +i_\eta R) \wedge ({\cal L}_\xi V+(d_\theta X +i_\xi R) \wedge V)  \nn
&=[{\cal L}_{\xi}, {\cal L}_{\eta} ]V +({\cal L}_{\xi} d_\theta Y)\wedge V
-({\cal L}_{\eta} d_\theta X)\wedge V +(({\cal L}_{\xi}i_\eta-{\cal L}_{\eta}i_\xi )R)\wedge V\nn
&={\cal L}_{[\xi, \eta]_\theta} V
+(d_\theta ({\cal L}_{\xi}Y-{\cal L}_{\eta}X))\wedge V
+i_{[\xi,\eta]_\theta}R -d_\theta i_\eta i_\xi R \nn
&={\cal L}_{[\xi, \eta]_\theta} V
+\left( d_\theta \left({\cal L}_{\xi}Y-{\cal L}_{\eta}X +\textstyle{\frac{1}{2}}d_\theta (i_X\eta -i_Y \xi ) 
-i_\eta i_\xi R\right) +i_{[\xi,\eta]_\theta}R\right) \wedge V \nn
&={\cal L}_{[X+\xi, Y+\eta]_R }V
\ena
where $[{\cal L}_{\xi}, {\cal L}_{\eta} ]={\cal L}_{[\xi, \eta]_\theta}$, 
$[d_\theta, {\cal L}_{\xi}]=0$ and $[{\cal L}_{\xi}, i_\eta]=i_{[\xi,\eta]_\theta}$ are used.
Note that
\bea
({\cal L}_{\xi}i_\eta-{\cal L}_{\eta}i_\xi )R
&=i_{[\xi,\eta]_\theta}R+i_\eta {\cal L}_{\xi}R-{\cal L}_{\eta}i_\xi R \nn
&=i_{[\xi,\eta]_\theta}R+i_\eta i_\xi d_\theta R -d_\theta i_\eta i_\xi R,
\ena
and $d_\theta R=0$.


\end{document}